\documentclass[aps,prd,a4paper,preprintnumbers,tightenlines,superscriptaddress,twocolumn,showpacs,final,nofootinbib,floatfix]{revtex4-1}
\usepackage{graphicx}
\usepackage{natbib}
\usepackage{amsmath,amssymb}
\usepackage{float}
\usepackage{dsfont}
\usepackage[toc,page]{appendix}
\usepackage{hyperref}

\newcommand{\mnras}{MNRAS}

\newcommand{\aj}{AJ}

\newcommand{\ha}{H$\alpha$}
\newcommand{\hb}{H$\beta$}
\newcommand{\gsim}{\;\mbox{\raisebox{-0.5ex}{$\stackrel{>}{\scriptstyle{\sim}}$}
}\;}
\newcommand{\lsim}{\;\mbox{\raisebox{-0.5ex}{$\stackrel{<}{\scriptstyle{\sim}}$}
}\;}
\newcommand{\rs}{{r_{\rm s}}}

\usepackage{color}

\begin{document}

\title{Astrophysical Tests of Modified Gravity: Stellar and Gaseous Rotation Curves in Dwarf Galaxies} 

\author{Vinu Vikram}
\affiliation{Argonne National Laboratory, 9700 South Cass Avenue, Lemont, IL 60439, USA}
\affiliation{Center for Particle Cosmology, Department of Physics and Astronomy, University of Pennsylvania 209 S. 33rd St., Philadelphia, PA 19104, USA}
\email{vvinuv@gmail.com }

\author{Jeremy Sakstein}
\affiliation{Center for Particle Cosmology, Department of Physics and Astronomy, University of Pennsylvania 209 S. 33rd St., Philadelphia, PA 19104, USA}
\affiliation{ DAMTP, Centre for Mathematical Sciences, University of Cambridge, Wilberforce Road, Cambridge CB3 0WA, UK}
 \affiliation{Perimeter Institute for Theoretical Physics, 31 Caroline St. N,Waterloo, ON, N2L 6B9, Canada}
\email{sakstein@physics.upenn.edu}

\author{Charles Davis}
\affiliation{Center for Particle Cosmology, Department of Physics and Astronomy, University of Pennsylvania 209 S. 33rd St., Philadelphia, PA 19104, USA}

\author{Andrew Neil}
\affiliation{Center for Particle Cosmology, Department of Physics and Astronomy, University of Pennsylvania 209 S. 33rd St., Philadelphia, PA 19104, USA}


\begin{abstract}
Chameleon theories of gravity predict that the gaseous component of isolated dwarf galaxies rotates with a faster velocity than
the stellar component. In this paper, we exploit this effect to obtain new constraints on the model parameters using the measured
rotation curves of six low surface brightness galaxies. For $f(R)$ theories, we rule out values of $f_{R0}>10^{-6}$. For more
general theories, we find that the constraints from Cepheid variable stars are currently more competitive than the bounds we
obtain here but we are able to rule out self-screening parameters $\chi_c>10^{-6}$ for fifth-force strengths (coupling of the scalar to matter)  as low as
$0.05$ the Newtonian force. This region of parameter space has hitherto been inaccessible to astrophysical probes. We
discuss the future prospects for improving these bounds.  
\end{abstract}

\maketitle
\section{Introduction}

Infrared modifications of general relativity (GR) have received recent attention as a possible candidate to explain the
acceleration of the universe (see \cite{Clifton:2011jh,Koyama:2015vza,Burrage:2016bwy,Burrage:2017qrf} and references therein for a
review). Amongst the plethora of proposed theories, those that include screening mechanisms---which give rise to novel features
on cosmological scales but hide any modifications of general relativity in our own solar system---are particularly well studied
due to their ability to satisfy local tests of general relativity without the need for fine-tuning. These fall into two
categories: the Vainshtein mechanism \cite{Vainshtein:1972sx}---which screens by suppressing scalar field gradients---and
those that screen by suppressing the scalar charge to mass ratio, which include the chameleon effect
\cite{Khoury03a,Khoury03b}, the symmetron mechanism
\cite{Hinterbichler:2010es} and the environment-dependent Damour-Polyakov effect \cite{Brax:2010gi}. This paper is concerned
with the latter class.

Recently, several authors \cite{hui09,changhui,jain11,bhuvjake2011,davis2011,jainvinu2012,Hui2012,Vikram2013,Sakstein13,Terukina:2013eqa,Wilcox:2015kna,Sakstein:2015oqa,Koyama:2015oma,Sakstein:2015zoa,Sakstein:2015aac,Sakstein:2017bws,Sakstein:2017pqi,Sakstein:2017xjx,Adhikari:2018izo} have
shown that astrophysical tests have the potential to probe parameter ranges inaccessible to laboratory experiments or cosmological
probes (for a review of these tests see\cite{jain-khoury10, jain11,Joyce:2014kja,Lombriser:2014dua,Burrage:2016bwy,Burrage:2017qrf}). These tests include the structure and
evolution of stars,
kinematical and morphological studies of dwarf galaxies, discrepancies between the dynamical and lensing masses of galaxy clusters, and the offset between compact objects and stellar components in dwarf
galaxies. In the first of a series of papers, \cite{cabre2012} compiled a screening map of the nearby universe, which identified
which galaxies are unscreened as a function of the model parameters. This was the first step towards making these (at the time)
theoretical tests possible. In the second paper, \cite{jainvinu2012} compared the observed Cepheid and tip of the red giant
branch distances to unscreened galaxies in the map and were able to place the strongest constraints to date. In the third paper
\cite{Vikram2013} attempted to perform the tests using the morphology and kinematics of the galaxies in the map but were unable
to place further constraints. This paper, the fourth in the series, is concerned with the final observational signature that has
not yet been tested observationally.

As mentioned above, chameleon-like theories screen by suppressing the scalar charge of an object relative to its mass and
\cite{bhuvjake2011} have used this feature to suggest a novel signature. Stars, being compact objects, are screened and hence
have zero scalar charge. This means their motion in modified theories of gravity is identical to that predicted by general
relativity. Conversely, diffuse gas is unscreened and feels the full fifth-force present due to the modifications. This means
that at fixed radius, the gaseous component of an unscreened galaxy should rotate with a higher velocity than the stellar
component. \cite{bhuvjake2011} estimate that this difference can be as high as 10-15 km/s depending on the galaxy's mass and the
model parameters.

\cite{Vikram2013} investigated the possibility of carrying out this test using currently available data but they were
unable to test the difference between the stellar and gaseous rotation curves. The main reason for this is the following:
Historically, the rotation curves of galaxies are measured using either \ha{} or the 21 cm line. \ha{} is produced by the
recombination of hot ionized gas around massive stars, called the Stromgren sphere, whilst the 21cm line is produced by neutral
hydrogen gas. Both of these lines probe the unscreened gaseous components of the galaxies. Hence, there is no screened component
with which to compare the measured rotation velocities and no quantitative statements could be made. This leaves the effect
described above an untested prediction. 

With high quality data, it would be possible to derive the stellar rotation curves from stellar absorption lines. In this paper we
do  this using data obtained for six low surface brightness galaxies by \cite{Pizzella2008} and attempt to constrain the
model parameters. One well-studied theory which utilizes the chameleon effect is $f(R)$ gravity \cite{Brax08}. The general class
of chameleon-like theories is parametrized by two numbers but $f(R)$ theories have one of these fixed, which makes them useful
prototype theories to explore. These theories are parametrized by the dimensionless number $f_{R0}$. More general chameleon (and similar) models have two parameters, $\chi_c$ and $\alpha$.

This paper is organized as follows: In section \ref{sec:theory} we give a brief introduction to chameleon theories of gravity and
describe the observational effect we are looking for. In section \ref{sec:data} we describe the data we will use and briefly
summarize the measurement of stellar and gaseous rotation curves. Section \ref{sec:systematics} describes the systematics involved
in the measurement and possible ways to correct it and section \ref{sec:analysis} describes the analysis of the rotation curves.
We discuss our results and draw our conclusions in sections \ref{sec:discussion} and \ref{sec:conclusion} respectively.

\section{Chameleon Screening and $f(R)$ Gravity }
\label{sec:theory}

In this section, we will briefly describe the screening mechanism and elucidate the model parameters.
Chameleon models include a new scalar degree of
freedom that couples to matter, giving rise to an additional gravitational strength force. Laboratory searches for fifth-forces
would generally constrain such theories to levels at which they become uninteresting but the chameleon mechanism acts to suppress
the fifth-force locally, thereby evading them. This is achieved
by arranging the field equations such that the scalar's mass in high-density environments is larger than the inverse micron scales
probed by current experiments. These experiments, therefore, leave a large region of parameter space
unconstrained. On larger---inter-galactic and cosmological---scales, the mass can be far smaller and $\mathcal{O}(1)$
fifth-forces can give rise to new and novel effects.

On small (astrophysical) scales, and when any relevant time-scale is small compared to the Hubble time, the entire class of
chameleon models can be parametrized by two dimensionless parameters $\alpha$ and $\chi_c$. Any over-dense spherical object of
radius $R$ embedded into a larger, under-dense medium will be characterized by a screening radius $r_{\rm s}$, which is not an
independent parameter, but is determined by $\chi_c$ in a manner to be made precise below. In the region interior to the screening
radius, there is no fifth-force, and the total force is the Newtonian one. Exterior to this region, the total force is given
by
\begin{equation}\label{eq:f5}
 F(r>r_{\rm s})=\frac{GM(r)}{r^2}\left[1+\alpha\left(1-\frac{M(r_{\rm s})}{M(r)}\right)\right]\quad r>r_{\rm s},
\end{equation}
where $M(r)$ is the mass enclosed by radius $r$. Note that when $r_{\rm s}=R$ the object is fully screened and the force is
simply the Newtonian one, whereas when $r_{\rm s}=0$ the object is fully unscreened and the strength of gravity is enhanced by a
factor of $(1+\alpha)$. $\alpha$ then parametrizes the strength of the chameleon force and is identically equal to $1/3$ in $f(R)$
models \cite{Brax08}.
The self-screening parameter $\chi_c$ determines how
efficient an object is at screening itself\footnote{Self-screening here refers to the fact that object may be screened by the
presence of massive neighbors. See \cite{hui09} for a discussion of this.} as a rule of thumb, if $\chi_c$ is smaller than the
surface Newtonian potential $GM/Rc^2$ the object is screened. If the converse is true, the object will be partially
unscreened. Very roughly, $\chi_c\sim H_0^2\lambda_{\rm C}^2 /c^2$ where $\lambda_{\rm C}$ is the Compton wavelength of the field
in the cosmological background. In $f(R)$ theories, one has $\chi_c=3/2f_{R0}$ \cite{hu07}
and therefore, have only one free parameter.
Currently, the strongest constraint on $f_{R0}$ was obtained by \cite{jainvinu2012} who find $f_{R0}\lsim
3\times10^{-7}$. Importantly, $\chi_c$ (or $f_{R0}$ for $f(R)$ theories) determines the screening radius through the implicit
relation
\begin{equation}\label{eq:screenrad}
 \int_\rs^\infty r\rho(r) dr=\frac{\chi_cc^2}{4\pi G}.
\end{equation}
If this has no solution, then $\rs=0$ and the object is unscreened. The derivation of this formula can be found in
\cite{hui09,davis2011,Sakstein13} but
here
we remark that the only approximation used to obtain it is that the mass of the field is negligible
compared with the inverse length scale in question. This is completely consistent with the approximation used to obtain equation
\ref{eq:f5} and is known to hold well inside stars and galaxies\footnote{see \cite{davis2011,Li12,Sakstein13} for a more detailed
discussion on this.}. Going beyond this approximation introduces a high degree of model-dependency and yields only small
corrections to the final results. In what follows we will use equation \ref{eq:screenrad} to determine the screening radius of
each galaxy in our sample assuming realistic density profiles to be specified below.

\section{Data and reduction}
\label{sec:data}
In this work we will use the rotation curves of six low surface brightness galaxies measured by \cite{Pizzella2008}. They
derived the rotation curves using both absorption and emission lines, which probe the rotation velocities of stars and gaseous
components respectively. Rotation curves are generated using data collected from the ESO Very Large
Telescope (VLT)-FORS2 instrument. The spectroscopic wavelength range of the spectra is between 4750 and 5800 \AA. This range
includes useful stellar absorption lines, such as the Mg\textit{b} triplet, and emission lines, such as \hb{} and [OIII]. The
rotation curve for the gaseous component was obtained by fitting Gaussians to these emission lines simultaneously. In addition to
the spectroscopic data, they acquired deep, high resolution photometric data for these galaxies. This was done using Gunn-z
filter, which probes the older stellar populations in the galaxy. Photometric data helps to identify the morphologies and generate
structure parameters for these galaxies. In table \ref{tab:basics} we show a few
relevant properties of these galaxies. 

 \begin{table}
\centering
 \begin{tabular}{ccccccc}
 Galaxy & D & i & B/T &$\phi_1$&$\phi_2$& $\phi_3$\\
\hline
ESO 4880490&37.5&67&0.21& 0&0&0\\
ESO 2060140&60.5&39& 0.04 &0&0&0\\
ESO 2340130 &60.9&69&0.66&$1.3\times10^{-6}$&$1.3\times10^{-6}$&$1.3\times10^{-6}$\\
ESO 4000370&60.1&50&0.03& 0&0&0\\
ESO 1860550 & 25 & 63 & 0.68 & $7.6\times10^{-8}$&$7.6\times10^{-8}$&
$1.7\times10^{-7}$\\
ESO 5340200&226.7&46&0.56& 0&0&0
\end{tabular}
\caption{Basic parameters of galaxies in our sample. (1) D: Distance to the galaxy (Mpc); (2) $i$ : inclination angle; (3) B/T : bulge to total ratio from
\cite{Pizzella2008}; (4) $\phi_1$, $\phi_2$, $\phi_3$: Newtonian potential of the environment for Compton scales of 1, 2 and 3
Mpc which corresponds to $f_{R0} \sim 1\times10^{-7}$, $f_{R0} \sim 4\times10^{-7}$ and $f_{R0} \sim 1\times10^{-6}$ respectively
\cite{sch09}.}
\label{tab:basics}
\end{table}

\subsection{Screening Level of Galaxies}
\label{sec:screenlevel}
It is essential for our test to know the screening level of a galaxy given a set of model parameters.
We check the screening level of a galaxy based on the procedure described in \cite{cabre2012}. {The screening map classifies the galaxies as either screened or unscreened based on two proxies that estimate whether they are self-screened due to their own Newtonian potential or environmentally screened due to the potential of their neighbors. In the former case, the criterion for self-screening is well understood: galaxies will be unscreened when the self-Newtonian potential $\Phi_{\rm N}^{\rm self}=GM/r_{\rm vir}c^2$---$r_{\rm vir}$ is the virial radius---is smaller than $\chi_c$($=3/2 f_{R0}$). This is estimated on a galaxy by galaxy basis using the relation $GM/r_{\rm vir} = v_{\rm c}^2$ where $v_{\rm c}$ is the peak circular velocity. The criterion for environmental screening is not so clear and there has been an intense effort in the N-body community aimed at finding reliable proxies for environmental screening \cite{Zhao:2011cu,zhao11}. The screening map uses the approximation that the level of environmental screening is set by the external Newtonian potential
\begin{equation}
\Phi_{\rm N}^{\rm ext}=\sum_{d<\lambda_{\rm C}+r_i}\frac{GM_i}{r_i},
\end{equation}
where $d_i$ is the distance to the galaxy with mass $M_i$ and virial radius $r_i$. The sum extends over galaxies inside the Compton wavelength ($\lambda_{\rm C}=1,\,3,\,10$ Mpc for $f_{R0}=10^{-7},\, 10^{-6},\,10^{-5}$). The motivation behind this is that the chameleon force is suppressed outside the Compton wavelength so that the fields sourced by any galaxies outside a sphere of radius $\lambda_{\rm C}$ will have attained their asymptotic values.    }

Based on the data described in \cite{cabre2012}, we estimate the Newtonian 
potential due to environment for all the six galaxies in our sample. It is found that, with the exception of galaxy ESO 2340130,
all other galaxies inhabit regions where the Newtonian potentials is low enough that they are not environmentally screened.
Therefore, it is possible to test chameleon theories if their self-Newtonian potential is less than the given value of $\chi_c$.
In table
\ref{tab:basics} we show the values of the Newtonian potential due to environment for different $f_{R0}$ values.

We derive the self-Newtonian potential of the galaxies using their stellar rotation curve. We are interested in probing
$\chi_c<10^{-6}$. At these low values, main-sequence stars are screened and hence have zero scalar charge. 
Therefore, they move according to general relativity and their rotation velocity traces the true mass of
the galaxy\footnote{In this context, the true mass of the galaxy is identical to its lensing mass.}. Also, we assume that the
galactic density profile is described by either an NFW or core-singular isothermal sphere (cSIS) model \cite{deblok10}: 

\begin{align}
\label{eq:nfw}\rho(r) &= \frac{\rho_0}{(r/r_0) (1+(r/r_0)^2)}&\textrm{NFW}\\
\rho(r) &= \frac{\rho_0}{1+(r/r_0)^2}&\textrm{cSIS},
\label{eq:cSIS} 
\end{align}
where $\rho_0$ and $r_0$ are the central densities and scale radii respectively. When determining the screening level of a
specific galaxy, we will use the empirically-fitted profiles in conjunction with equation (\ref{eq:screenrad}) to find the
screening radius. The $\infty$ in the upper limit is really a proxy for the radius at which the field reaches its cosmic value.
The NFW profile (equation (\ref{eq:nfw})) falls off sufficiently quickly at large radii that this may be performed exactly but the
cSIS
profile (equation (\ref{eq:cSIS})) has a slower fall-off and is unphysical at large radii. For this reason, we integrate to
$R_{\rm 200}$
in order to determine the screening radius. One should really integrate to a few Compton wavelengths out from this, however this
will introduce some model-dependency. Integrating to $R_{200}$ is completely consistent with the approximation that the mass is
negligible, and we have checked that the results are robust to changing the upper limits to a few times $R_{\rm 200}$. 

It should be noted that the dominant dark matter component of the galaxy could be unscreened so that it may deviate from the
standard NFW or cSIS profiles \cite{lombriser12}. We do not consider this effect in this study. We fit the stellar rotation curve
to derive the scale radii and central densities of for each profile. These fitted parameters are then used to estimate the
Newtonian potential of the galaxies. Based on the uncertainties in the fitted parameters, we found that cSIS gives a better fit to
all of the rotation curves except for ESO 5340200. In table \ref{tab:fitcSIS} we show the fitted parameters for the cSIS and NFW
models. Based on the environmental and self-Newtonian potentials, we estimate the value of $f_{R0}$ above which each galaxy
becomes unscreened. This implies that each galaxy allows us to probe different values of $f_{R0}$. These values are
also shown in table \ref{tab:fitcSIS}. 

{Finally, we note that there may be uncertainties in the environmental screening level due to \emph{missing mass} \cite{Li:2012by,Desmond:2017ctk}. In particular, N-body simulations have shown that density perturbations on scales smaller than $\sim 10$ Mpc could be enhanced by factors of $50$--$100$\% due to dark matter that does not include any visible host galaxies or that hosts galaxies too faint to be resolved in the 2M++ survey. These would still contribute to the environmental value of the Newtonian potential and could potentially enhance the screening level by $5$--$10\%$. Unfortunately, missing mass is not accounted for in the screening map of \cite{cabre2012} so we are unable to estimate its effects on our galaxies. One could reduce the uncertainty associated with missing mass using forward Bayesian modeling applied to large sample sizes as was done recently by reference \cite{Desmond:2018euk}.}

\subsection{Spectral Lines Used in the Analysis}
It is important to be sure that we are probing both the screened and unscreened components. This is achieved by our choice of
absorption and emission lines. The Mg\textit{b} triplet lines are due to absorption in the atmospheres of G- and K-type stars
\cite{boselli2012panchromatic}. These are low mass ($0.6$--$1.2M_\odot$) main-sequence stars with temperatures between $4000$
and $7000$ K. These stars have Newtonian potentials $\Phi_{\rm N}\sim\mathcal{O}(10^{-6})$ and are hence self-screened when
$\chi_c\lsim 10^{-6}$. We can therefore use the rotation curve measured using these lines as a tracer of the screened component
of the galaxy. The three absorption lines used by \cite{Pizzella2008} have wavelengths 5164, 5173 and 5184 \AA. It should be
noted that giant stars can also contribute to the Mg\textit{b} absorption lines but this is small enough to ignore in this study
\cite{conroy13}. On the other hand, [OIII]
emission lines result from forbidden transitions from doubly ionized oxygen in a metastable state. This line only exists at
extremely low densities. At higher densities, inelastic collisions relax the system to a stable state without the emission of photon. These lines therefore probe the unscreened gaseous component of the galaxies. 

Before going further, it is instructive to pause and comment on the range of $\chi_c$ that can be probed with our rotation
curve analysis. As remarked above, the main-sequence stars used to calculate the stellar rotation curve have Newtonian
potentials of order $10^{-6}$ and are hence unscreened when $\chi_c\gsim 5\times10^{-6}$. When this is the case, the stars probed
by the Mg\textit{b} absorption lines rotate at the same speed as the gaseous component and there is no offset between the two
components. Therefore, the offset of the two rotation curves probes only a narrow range in $\chi_c$.
The lower limit is set by the Newtonian potential of the galaxy and/or the environment (typically
$\mathcal{O}(10^{-8})$--$\mathcal{O}(10^{-7})$ for the galaxies in our sample) and the upper limit is set by the value of $\chi_c$
above which
G- and K-type stars unscreened\footnote{In practice, most stars will be unscreened when $\chi_c\gsim5\times10^{-6}$. Main-sequence
stars obey a well-known mass-radius relation $R\propto M^{\nu}$, where $\nu$ lies in the range $0.2$--$0.8$ depending on the
stellar mass. Stars bluer than those probed by the Mg\textit{b} triplet are more massive and hence have smaller Newtonian
potentials. Redder stars have $\nu\approx0.2$ appropriate for the PPI chain and so the Newtonian potential is a very weak function
of mass for these stars.}. 

\begin{table}
\begin{center}
 \begin{tabular}{ccccccc}
 Galaxy & \multicolumn{2}{c}{cSIS}& $f_{R0}$ \\
 & $r_0$ & $\rho_0$  &  \\
\hline
ESO 4880490 & $2.0 \times 10^{3}$ & $4.0 \times 10^{-2}$ &$2.34\times10^{-7}$\\
ESO 2060140  & $3.1 \times 10^{3}$ & $5.1 \times 10^{-2}$ & $7.39\times10^{-7}$\\
ESO 2340130 & $1.1 \times 10^{3}$ & $3.9 \times 10^{-1}$ & $9.04\times10^{-7}$\\
ESO 4000370 & $7.2 \times 10^{3}$ & $3.7 \times 10^{-2}$ &$2.77\times10^{-6}$ \\
ESO 1860550  & $1.5 \times 10^{3}$ & $3.5 \times 10^{-1}$ & $1.49\times10^{-6}$  \\
\hline
 Galaxy & \multicolumn{2}{c}{NFW}& $f_{R0}$ \\
 & $r_0$ & $\rho_0$  &  \\
 \hline
ESO 5340200 & $3.4 \times 10^{3}$ & $4.2 \times 10^{-1}$ &$1.94\times10^{-6}$
\end{tabular}
\end{center}
\caption{The best fit cSIS and NFW parameters for the stellar rotation curves.  Columns: $\rho_0$ and $r_0$ are defined as in
Eqns. \ref{eq:nfw} \& \ref{eq:cSIS}. These have unit $M_\odot$ pc$^{-3}$
and parsec. $f_{R0}$ is the value above which the galaxy is completely unscreened.}
\label{tab:fitcSIS}
\end{table}

\section{Systematic corrections to the rotation curves}
\label{sec:systematics}
It is important to understand different systematic uncertainties associated with rotation curves based on stellar absorption lines
and gaseous emission lines. In this section we
describe the two major systematics in the rotation curve estimates which could possibly mimic a modified gravity signal. Gaseous
rotation curves are known to be affected by non-circular motions. This could be due to the presence of star forming regions and
other morphological components such as bars and spiral structures \cite{oh08}. Based on a sample of 19
galaxies, \cite{trachternach08} have shown that the non-circular velocity is $\approx 7$ km/s for galaxies with similar
magnitudes and rotational velocities to those in our sample. 
This systematic component in the gaseous rotation curve can potentially mimic a modified gravity signal as it makes the gaseous
rotation appear faster than it truly is. In our analysis we subtract the average velocity found by \cite{trachternach08} from all
the gaseous rotation curves as follows:
\begin{equation}
V_{\rm gas, cor}^2 = V_{\rm gas, obs}^2 - V^2_{\rm sys}
\end{equation}
where $V_{\rm gas, obs}$, $V_{\rm gas, cor}$, $V_{\rm sys}$ are the observed, corrected rotation velocities and the non-circular
velocity respectively.

In addition to the above systematic, asymmetric drift could potentially make the stellar 
rotation appear slower than the true underlying circular velocity, which could introduce pseudo signal into the data. We
attempt to correct the observed rotation velocity based on the observed velocity dispersion and Jeans equations using the
following assumptions: According to \cite{binney87, hinz01} if:  

\begin{enumerate}
\item the stellar component follows an exponential density profile, 
\item $\sigma_\phi^2/\sigma_R^2 \approx 0.5$, where $\sigma_\phi$ and $\sigma_R$ are the velocity dispersions along the azimuthal
and radial directions (in cylindrical coordinates) respectively,
\item $v_z v_R \approx 0$, i.e. the product of the velocities along radial and perpendicular ($z$ direction) to the disk is
small and
\item $\partial\ln\sigma^2_R/\partial\ln R \approx 0$, i.e the gradient of radial velocity dispersion is small
\end{enumerate}
then the corrected circular velocity can be written as 
\begin{equation}
v_c^2 = v_\phi^2 + \sigma_\phi^2 \left(2 \frac{R}{R_{\rm exp}} - 1 \right),
\label{eq:asym-drift}
\end{equation}
where $R_{\rm exp}$ is the scale radius of the exponential disk. \cite{Pizzella2008} shows that the galaxies in our sample can be
modelled with one or two exponential profiles but three of these galaxies are dominated by the inner exponential bulge
component. This is evident from the bulge-to-total light estimated by \cite{Pizzella2008} (see table \ref{tab:basics}). We
exclude the central bulge dominated region of the galaxies from the analysis to reduce any non-circular contamination from bulge. It should be noted that in equation
(\ref{eq:asym-drift}) we use the average value of the measured velocity dispersion outside the bulge to estimate the correction.

In the latter part of the paper we will see that asymmetric drift correction leads to larger
values for stellar rotation curves. This implies that without the asymmetric drift correction, the stellar rotation curve
appears systematically lower and can be misidentified as modified gravity signal when compared with the gaseous rotation curve.

\section{Analysis of the Rotation Curves}
\label{sec:analysis}
Once the systematics are accounted for, any difference between the stellar and gaseous rotation velocities is a probe of modified
gravity. The stellar and gaseous curves are measured at slightly different radii from the centre of the galaxy. Therefore, in
order to estimate the difference between them we need to interpolate one or the other to a common radius. 
Since the gaseous rotation curve is measured at many more finely spaced points than the stellar rotation curve, it is better to 
interpolate the gaseous rotation curves and leave the stellar curves as measured. Finally, the average values of the gaseous and
stellar rotational velocities of both sides of the galaxy is found. It is also possible to take
the averages of the rotation curves without interpolation, however this average depends on the locations of the measured
points. This means that if there are many measured points in the inner regions, the average can be biased towards a lower
rotational
velocity. 

The gaseous rotation curve can be measured out to larger radii than the stellar curve. Therefore, in order to make a fair
comparison with the stellar rotation curve, we need to truncate the gaseous rotation curve at radii larger than the maximum
radius where the stellar curve is measured. This ensures that we are comparing the rotation velocities in the same
region of the galaxy. In addition to this upper limit on the rotation curve radius, we also need to set a lower limit, which helps
to avoid the bulge dominated central region of the galaxy. This means that we consider only those velocity points measured at
radii larger than the bulge scale radius $R_e$. We check whether the final results depend on this choice of radius by repeating
the analysis with a range of different lower limits, ranging from $0.5R_e$ to $1.75R_e$. In all cases we find that the final
result does not depend on the choice of lower limit, except for the galaxy ESO 1860550. ESO 1860550 shows an increase in the
difference between the gaseous and stellar rotation curves between 0.5 and 1.5 times the bulge scale radius. The difference in
velocity curves increases from $-25 \pm 7$ km/s at $0.5R_e$ to $-60 \pm 9$ km/s at $1.75R_e$. This galaxy is bulge dominated
(B/T= 0.68, see table \ref{tab:basics}) with a concentrated bulge component (Sersic index $n > 2$) compared with the rest of the
galaxies, and therefore this increase may be associated with the large velocity dispersion in the bulge. Later it can be seen that ignoring larger number of observations from the inner part of the rotation curve only strengthen our results.

We give the inverse variance weighted average values of the stellar and gaseous rotation curves in table \ref{tab:delta_v}; we
assume that the measured points are uncorrelated. Next to the average value we show the statistical error. 
In addition to the statistical error, there could be some intrinsic scatter associated with the average rotation velocity.
This is due to the fact that the observed rotation curve is not very smooth. This scatter could be introduced by different
components such as star forming regions, dust, and other morphological components of the galaxies. We estimate the intrinsic
scatter from the residual rotation curve by subtracting the best fit model from the observed rotation curve. This is done
separately for both the stellar and gaseous rotation curves. We quantify the intrinsic errors as follows:
\begin{equation}
\sigma_{\rm int} = \sigma_v / \sqrt{N}
\end{equation} 
where $\sigma_v$ is the scatter in the residual rotation velocity curve and $N$ is the number of measured velocity data points.
The intrinsic scatter is shown in table \ref{tab:delta_v} after the statistical error. It can be seen that the intrinsic error
dominates over the statistical error for most of the galaxies. 

One of the most important sources of error in all astrophysical tests of gravity is the uncertainty due to non-gravitational
astrophysical processes. This means that the difference between the
measured stellar and gaseous rotation velocities could be due to astrophysical scatter instead of modified gravity effects.
Earlier papers in the series estimated this uncertainty from a sample of screened galaxies and subtracted it statistically from
the unscreened sample. We apply the same technique in this paper. The number of objects in the current sample is very small and
therefore it is not possible to estimate the astrophysical scatter from a screened sample. We therefore take the scatter in
the measured velocity difference of all six galaxies as the astrophysical scatter ($\sigma_{\rm ap}$). In this process we assume
that the scatter measured from all galaxies is closer to the underline true value (i.e. the one measured from screened galaxies
only) and that the true scatter can be measured from six galaxies.  
It is possible that the $\sigma_{\rm ap}$ could be overestimated by this process and therefore our conclusions may be conservative. Finally, similar to the earlier papers we assume that $\sigma_{\rm ap}$ scales with the mass of the galaxy. $\sigma_{\rm ap}^i$ of $i^{\textrm{th}}$ galaxy is estimated as 
\begin{equation}
\sigma_{\rm ap}^i = \frac{v_i}{N}\sum{\frac{\delta_v}{v}}
\end{equation}
where $v_i$ is the average of the stellar and gaseous velocities of the $i^{\textrm{th}}$ galaxy, $\delta_v$ is the uncorrected
velocity difference between the gas and stars; the sum runs over all six galaxies. We show $\sigma_{\rm ap}$ in table
\ref{tab:delta_v} as the third source of error.

{The scatter in table \ref{tab:delta_v_uncorr} is commensurate with the scatter seen in N-body simulations, although it should be noted that observed variations in the rotation curves of dwarf galaxies is typically larger than this \cite{Oman:2015xda}. This implies that there may be unknown systematics not accounted for in our astrophysical scatter. Another potential uncertainty is a potential deviation from the NFW profile in the central regions of the galaxy. Dwarf galaxies, being dark matter dominated are not so susceptible to the effects of baryons at the center and, besides, we exclude the central bulge from our analysis since contributions from the non-circular velocity to the rotation curve may be important there. Nonetheless, the simple NFW profile ignores dark matter substructure, which can contribute as much as 10\% of the Halo's mass \cite{Hellwing:2015upa}. This could potentially increase the galaxy's level of self-screening. Finally, high resolution N-body simulations have shown that the halo mass function can be enhanced by up to 20\% even in cases where $f_{R0}\sim 10^{-6}$ so one may expect more self-screened galaxies. Our galaxies are selected from the screening map of \cite{cabre2012}, which uses proxies for the self-screening calibrated on the N-body simulations of \cite{Zhao:2011cu,zhao11}. This suggests that our classification of galaxies as unscreened is robust to enhancements in the halo mass function, although there may be small uncertainties in the proxies themselves if the density enhancement was not identified in the simulations of \cite{Zhao:2011cu,zhao11}. A quantitative accounting of these effects is clearly outside the scope of the present work due to the limitations of the screening map but future efforts aimed at constructing more comprehensive maps could elucidate how the features mentioned here impact the screening status of individual galaxies \cite{Desmond:2017ctk}.   }

\begin{figure*}
  \begin{center}
     \includegraphics[scale=0.6]{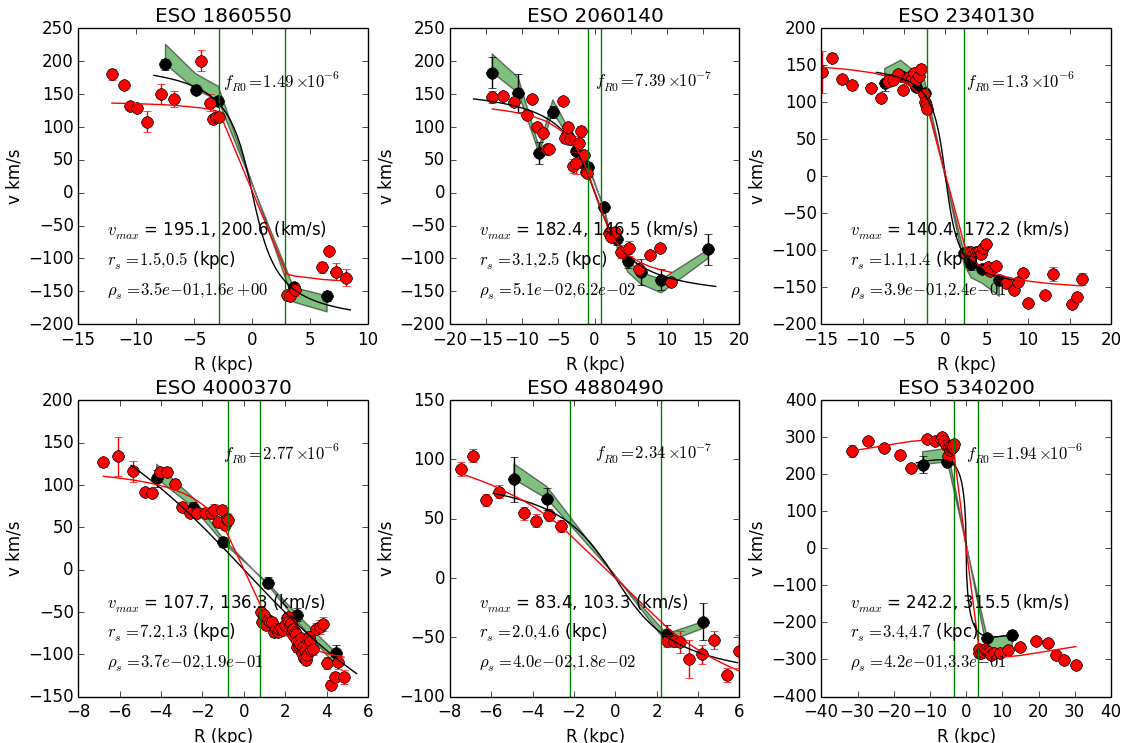}  
  \end{center}
  \caption{The rotation curves of the six galaxies in our sample. The black and red points show the stellar and gaseous rotation
curves respectively. The green shaded region shows the $f(R)$ prediction for the rotation velocity of the gaseous component
assuming that the galaxy is fully unscreened. The value of $f_{R0}$ which makes each galaxy fully unscreened is given in top right
side of each panel. We ignore the possible radial dependence of fifth force while estimating the prediction. 
The vertical green lines show the bulge
scale radius, below which we ignore the rotation measurements from the analysis and for fitting the density profile. In all but
the lower right panel, the black and red curves show the fitted cSIS profiles for the stellar and gaseous rotation curves
respectively. The lower right panel shows the fitted NFW profile. It can be seen that the fitted values for the gaseous and
stellar rotation curves differ slightly due to the fact that there are only a few points for the stellar curves, all of which are
at very small distances from the galactic centre. $v_{\rm max}$ is the the maximum measured rotation velocities for the
stellar and gaseous components respectively. We also show the fitted values of $r_0$ and $\rho_0$ for the relevant profiles
(cSIS for the first five panels and NFW for the lower right) for the stellar and gaseous rotation curves respectively. Details of
the analysis can be found in the text.}
\label{fig:rc}
\end{figure*}

\section{Results \& Discussion}
\label{sec:discussion}

\begin{table}
\begin{center}
 \begin{tabular}{ccc}
 Galaxy & $v_{\rm gas}$ (km/s) & $v_{\rm star}$ (km/s) \\
 \hline
 ESO 4880490 & 56.56 $\pm$ 2.77 $\pm$ 1.93 & 48.86 $\pm$ 5.37 $\pm$ 6.04 \\
 ESO 2060140 & 88.42 $\pm$ 1.65 $\pm$ 3.59 & 65.77 $\pm$ 2.39 $\pm$ 6.54 \\
 ESO 2340130 & 114.16 $\pm$ 1.87 $\pm$ 1.69 & 110.75 $\pm$ 2.24 $\pm$ 1.63 \\
 ESO 4000370 & 83.51 $\pm$ 2.24 $\pm$ 1.42 & 57.92 $\pm$ 2.99 $\pm$ 7.07 \\
ESO 1860550 & 119.91 $\pm$ 3.57 $\pm$ 5.32 & 137.79 $\pm$ 3.15 $\pm$ 2.69 \\
ESO 5340200 & 272.01 $\pm$ 3.65 $\pm$ 2.82 & 239.45 $\pm$ 4.56 $\pm$ 6.15 \\
\end{tabular}
\end{center}
\caption{The measured stellar and gaseous velocities. These values are not corrected for asymmetric drift and
non-circular motions.}
\label{tab:delta_v_uncorr}
\end{table}

\begin{table*}
\begin{center}
 \begin{tabular}{cccccc}
 Galaxy & $v_{\rm gas}$  & $v_{\rm star}$  & $\delta_v$  & $\delta_v^{\rm predicted}$ & Rejection $\sigma$\\
 \hline
 ESO 4880490 & 56.12 $\pm$ 2.77 $\pm$ 1.93 & 55.34 $\pm$ 5.37 $\pm$ 6.04 & 0.79 $\pm$ 6.04 $\pm$ 6.34 $\pm$ 5.85 (10.36) & 8.56 &
-0.8 (-0.9)\\
 ESO 2060140 & 88.11 $\pm$ 1.65 $\pm$ 3.59 & 65.39 $\pm$ 2.39 $\pm$ 6.54 & 22.72 $\pm$ 2.91 $\pm$ 7.46 $\pm$ 8.56 (9.50) & 10.12 &
1.3 (1.6)\\
 ESO 2340130 & 113.94 $\pm$ 1.87 $\pm$ 1.69 & 123.64 $\pm$ 2.24 $\pm$ 1.63 & -9.70 $\pm$ 2.92 $\pm$ 2.35 $\pm$ 12.49 (13.15) &
19.13 & -2.2 (-7.7) \\
 ESO 4000370 & 83.19 $\pm$ 2.24 $\pm$ 1.42 & 58.42 $\pm$ 2.99 $\pm$ 7.07 & 24.78 $\pm$ 3.74 $\pm$ 7.21 $\pm$ 7.85 (9.47) & 9.04 &
1.7 (1.9)\\
ESO 1860550 & 119.69 $\pm$ 3.57 $\pm$ 5.32 & 158.83 $\pm$ 3.15 $\pm$ 2.69 & -39.14 $\pm$ 4.76 $\pm$ 5.96 $\pm$ 14.31 (15.81) &
24.57 & -4.0 (-8.4)\\
ESO 5340200 & 271.91 $\pm$ 3.65 $\pm$ 2.82 & 236.40 $\pm$ 4.56 $\pm$ 6.15 & 35.51 $\pm$ 5.84 $\pm$ 6.77 $\pm$ 28.40 (29.58) &
36.57 & 0.0 (-0.1)\\

\end{tabular}
\end{center}
\caption{The stellar and gaseous rotation velocities for each galaxy in our sample after correcting for systematics. $v_{\rm
gas}$ and $v_{\rm star}$ are the rotation velocities of the gaseous and stellar components in the range outside the
bulge component of the galaxy respectively.  $\delta_v$ = $v_{\rm gas} - v_{\rm star}$ and $\delta_v^{\rm predicted}$ is the
$f(R)$ prediction for this difference. All the velocities are in units of km/s. The first part of the error is statistical and the
second part is the intrinsic scatter in the rotation curve. The third error in $\delta_v$ shows the astrophysical scatter and the
value in the bracket shows the effective error including all three sources. The rejection $\sigma$ is the confidence
with which we can reject predictions from $f(R)$ gravity. The first part of this column shows the significance after including all
sources of error in the analysis and the value in brackets shows the significance without including the astrophysical scatter.
A negative (positive) rejection $\sigma$ implies that stars rotate faster (slower) than the gas.}
\label{tab:delta_v}
\end{table*}

In tables \ref{tab:delta_v_uncorr} and \ref{tab:delta_v} we show the gaseous and stellar rotation velocities before and
after systematic corrections along with statistical error, intrinsic and astrophysical scatters in the measurements. Table \ref{tab:delta_v} 
shows the observed difference between the stellar and gaseous velocities and the $f(R)$ prediction. The $f(R)$ prediction is valid for the range of  $f_{R0}$  which makes the galaxies unscreened; this is given in Table \ref{tab:fitcSIS}. 
Therefore, $G_{\rm eff} = 4/3G$ is used for all galaxies to get the prediction. Tables \ref{tab:delta_v_uncorr} and \ref{tab:delta_v} show that the trend found
in the observed velocity differences are qualitatively similar for all of the galaxies both before and after the systematic
correction. We therefore use the systematic corrected values from here on. The final column
in these tables shows the statistical significance ($\sigma$) with which we can reject the predicted difference between the
gaseous and
stellar rotation curves based on the observed value and is defined as
\begin{equation}
\sigma = (\delta_v^{\rm measured} - \delta_v^{\rm predicted})/\rm{Measured~error} 
\end{equation}
 i.e. $\sigma = 0$ implies that the predicted and measured values agree perfectly and negative (positive) $\sigma$ imply that
stars rotate faster (slower) than the gaseous components. $f(R)$ theories,
and, indeed, more general chameleon-like theories predict that $\Delta G/G>0$ for the gaseous component and so cannot explain
any observations of stars rotating faster than the gas. Additionally, $f(R)$ theory cannot explain a faster gaseous rotation than
the predicted values given in table \ref{tab:delta_v}\footnote{More general chameleon models can due to the freedom in the
additional parameter $\alpha$.}.
Below, we comment on individual galaxies. The list is ordered in such a way that the first galaxy probes the smallest value of
$f_{R0}$ and last probes the largest. Also, we use the rejection sigma estimated after including astrophysical scatter
for the following discussion. Here, we concentrate on $f(R)$ theories in order to focus on a concrete model and extend the
analysis to more general models below.
\begin{enumerate}

\item \textbf{ESO 4880490:}  Both GR and $f(R)$ theories agree
with the measured values within $1 \sigma$. However, this galaxy is self screened for $f_{R0} < 2.34 \times 10^{-7}$ and conclusions about $f(R)$ theories are applicable only for theories with $f_{R0}$ larger than this. The intrinsic scatter in the rotation curve, which is common to dwarf galaxies, and
astrophysical scatter are the major limiting factors which prevents us from drawing a stronger conclusion. This galaxy posses a
small exponential bulge \footnote{Exponential bulges are different from classical bulges. The latter, found in elliptical
galaxies, follow de Vaucouleurs' light profile and are supported by random velocity. On the other hand, exponential bulges are
commonly found in disk galaxies. They follow exponential light profiles and are supported mostly by rotation.}.

\item \textbf{ESO 2060140:} $f(R)$ gravity agrees with the measurements more than GR. From table \ref{tab:fitcSIS} it can be seen that this galaxy becomes  self screened for $f_{R0} < 7.39 \times 10^{-7}$. 
This galaxy is a pure disk system with B/T = 0.04, which makes the galaxy a good candidate for testing modified gravity. 

\item \textbf{ESO 2340130:} This galaxy rejects $f(R)$ theories with with $\sim 2 \sigma$ and agrees with GR.  This galaxy is screened by  environment for $f_{R0} < 1.3\times 10^{-6}$. 

\item \textbf{ESO 4000370:} This galaxy agrees with the $f(R)$ prediction (within $\sim 1 \sigma$) better than the GR
prediction (which is $> 2\sigma$ away). This galaxy is self screened for $f_{R0} < 2.77 \times 10^{-6}$ and therefore, the
rejection is applicable only to models with $f_{R0}$ larger than this.

\item \textbf{ESO 1860550:} This galaxy rejects $f(R)$ with more than $3 \sigma$ confidence. This galaxy is self screened for $f_{R0} < 1.49 \times 10^{-6}$. As described in Section \ref{sec:analysis} removing more measured points from the inner part of  rotation curves strengthen the significance of rejection.  

\item \textbf{ESO 5340200:} This galaxy agrees with the predictions of $f(R)$ and GR within $1 \sigma$ but may
not be able to constrain theories with $f_{R0} < 1 \times 10^{-6}$ due to self-screening.
\end{enumerate}

It is important to note that the different screening levels of each galaxy mandate that we consider each separately
when calculating the confidence with which we can reject modified gravity. This is not the case with GR since the equivalence
principle is obeyed. In this case, one should take the average over all six galaxies and so the rejection $\sigma$ for GR stated
above should be taken equivocally; they are for comparison purposes only. Averaging over all six galaxies, we find
$\langle\delta_v/v\rangle=0.07 \pm 0.13$ and so the data is perfectly consistent with GR. A larger sample of unscreened
galaxies would allow us to perform a similar analysis for modified gravity; such a comparison is meaningless for our current
sample. It is also interesting to note that the galaxies ESO 2060140 and ESO 4000370 have similar rotation velocities and show a similar
trend in the stellar and gaseous rotation curves.

\begin{figure}
  \begin{center}
     \includegraphics[scale=0.5]{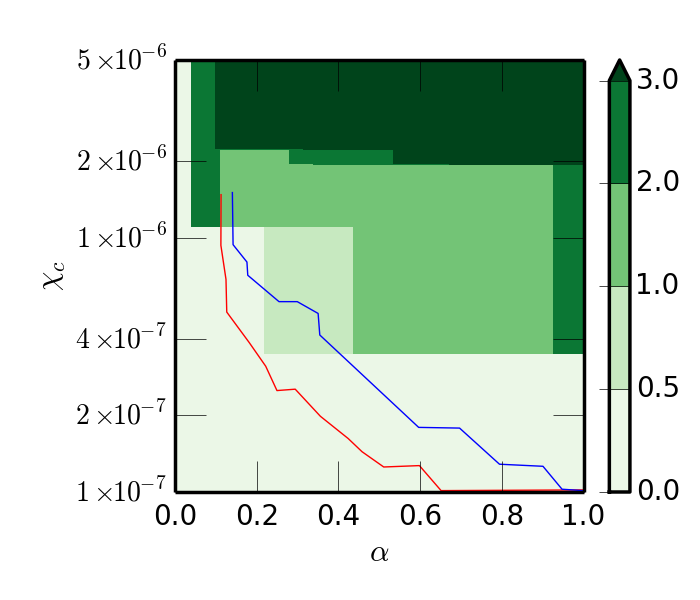}
     \end{center}
  \caption{Green contours show the limits obtained on the two parameters ($\chi_c$ \& $\alpha$) of chameleon theories. We show contours  of $0.5 \sigma$, $1 \sigma$, $2 \sigma$,   $3 \sigma$  where $\sigma$ represents the confidence with which the corresponding parameter values can be rejected. We also show the $1 \sigma$ and $2 \sigma$ contours from \cite{jainvinu2012}. It can be seen that the previous study give better constraints for $\alpha > 0.1$. This work is able to give upper limits on $f_{R0}$ value for $\alpha < 0.1$.}
\label{fig:constraints}
\end{figure}

Having elucidated the galaxies varying degrees of usefulness for probing modified gravity, we extend the analysis to general
chameleon models to constrain the parameters $\chi_c$ and $\alpha$. We consider $\alpha$ values between 0.05 and 1. Within this
range, the Compton wavelength does not vary more than a factor of a few compared with the value at $\alpha=1/3$\footnote{This
is important for addressing the environmental screening status of the galaxies, which is calibrated on $f(R)$ models
\cite{cabre2012}.}.  
Figure \ref{fig:constraints} shows the combined constraints obtained for $\alpha$ and $\chi_c$ values using all the six galaxies.
We show contours of $0.5 \sigma$, $1 \sigma$, $2 \sigma$, $3 \sigma$  where $\sigma$ represents the confidence with which we
can reject the corresponding parameter range. We also show the previous constraints obtained by
\cite{jainvinu2012}. It can be seen that the constraints obtained from Cepheid observations are stronger compared
to this analysis, especially for $\alpha > 0.1$. For values of $\alpha < 0.1$ we find that rotation curves give a better upper
bound on $\chi_c$.

\section{Summary}
\label{sec:conclusion}

Chameleon theories of gravity predict large differences between the stellar and gaseous rotation curves of isolated dwarf
galaxies. Since stars are screened, they are expected to move according to GR whilst the unscreened gaseous component
is predicted to rotate with a larger velocity. In this paper, we have used this effect to constrain the model parameters using
observations of the rotation curves of six low mass galaxies.

The main results are summarised in figure \ref{fig:constraints}, where we show the upper limits on the parameters, $\chi_c$
and $\alpha$. For $f(R)$ theory, i.e. when $\alpha=1/3$, we found that $f_{R0}>1.5\times10^{-6}$ can be ruled out with more than
$3 \sigma$ confidence. For a general chameleon theory $\chi_c > 10^{-6}$ can be ruled out with 95\% confidence down to $\alpha
\approx 0.05$. This is a new and strong result, and it arises because the measured difference in velocity has the opposite sign
from the chameleon predictions for some of the galaxies. This region of parameter space was previously unexplored due to the
small signals predicted when $\alpha\ll1$. 

Other studies have constrained chameleon theories as well and here we discuss how our new constraints compare with
previous bounds obtained using astrophysical tests\footnote{See \cite{Sakstein13,bhuv2013,Joyce:2014kja} for a review of
astrophysical tests.}. \cite{Vikram2013} attempt to constrain $f(R)$ gravity based on several morphological and kinematical
signatures using dwarf galaxies. Some of those tests indicate that $f_{R0} < 4\times10^{-7}$ but they show that better data is
required to get significant constraints on the model parameters. Studies based on distance indicators
\cite{jainvinu2012} were able to rule out $f_{R0}>4\times10^{-7}$ with more than 95\% confidence. Our constraints at low $\alpha$
are an improvement over these constraints but at larger values the Cepheid bounds are stronger. It is worth noting that the novel
features that arise in Cepheids, which are partially screened, are due to the increase in the strength of gravity and so they are
not sensitive to small values of $\alpha$. The test described here works when the galaxy is fully unscreened which is why we are
able to probe into the regime $\alpha\lsim0.2$.
Given that these are the early stages of testing modified gravity, it is valuable to use many different probes that rely on
different physical systems, such as pulsating stars and disk galaxies. 

As shown in table \ref{tab:delta_v}, the astrophysical scatter in the rotation curve  is one of the major limiting factors
in this study. Such statistical uncertainties can be reduced by combining observations of several galaxies. For example, it can be
seen that galaxies such as ESO 4880490 probe $f_{R0}$ down to $2\times 10^{-7}$ but there is a large
statistical uncertainty associated with the measurement. Based on the measured uncertainty, it can be seen that
$\approx 25$ similar low mass galaxies are enough to rule out $f_{R0} = 2\times10^{-7}$ with more than $4\sigma$ confidence.
Future targeted observations of high signal-to-noise stellar rotation curves of isolated dwarf systems in the local Universe may
serve that purpose.

One of the major assumptions in the above forecast is that the observed rotation velocity curves are free from systematics.
Therefore, proper corrections for systematics such as asymmetric drift and non-rotational velocities are needed. Finally, it
should be noted that several independent constraints on $f(R)$ parameters exist in the literature based on cosmological and lab
experiments \cite{lombriser12,bhuv2013}. In this paper we limit our discussion only to astrophysical tests, which are emerging as
a novel way of testing gravity. 

\section*{Acknowledgements}
We are grateful to Bhuvnesh Jain
for in-depth discussions and helpful suggestions. We thank Matthew Walker for many helpful suggestions.
This work benefited from discussions with Lasha Berezhiani, Cullen Blake, Yi-Zen Chu, Joseph Clampitt, Mike Jarvis, Justin Khoury, Kazuya
Koyama, Baojiu Li, Adam Lidz, Claudia Maraston, Alan Meert, Fabian Schmidt and Mitchell Struble. JS is supported by funds provided
to the Center for Particle Cosmology by the University of Pennsylvania. Research at Perimeter
Institute is supported by the Government of Canada through Industry Canada and by the
Province of Ontario through the Ministry of Economic Development \& Innovation.

\bibliography{rotation_curve_jcap}

\end{document}